\documentclass[apj]{emulateapj}
\usepackage{graphicx}

\newcommand{\lsun}{L$_{\odot}$}
\newcommand{\sfr}{M$_{\odot}$ yr$^{-1}$}

\newcommand{\oiii}{[O{\sc iii}]}
\newcommand{\cii}{[C{\sc ii}]}
\newcommand{\lcii}{L$_{\rm[CII]}$}
\newcommand{\lya}{Ly$\alpha$}
\newcommand{\acii}{A$_{\rm [CII]}$}
\newcommand{\bcii}{B$_{\rm [CII]}$}

\received{Dec 4, 2017}
\revised{}
\accepted{\today}

\shorttitle{\ sample article}
\shortauthors{Carniani et al.}

\begin{document}

\title{ALMA detection of extended [CII] emission in Himiko at z=6.6}


\author{S. Carniani\altaffilmark{1,2}, 
R. Maiolino\altaffilmark{1,2},
R. Smit\altaffilmark{1,2},
and R. Amor\'in\altaffilmark{1,2}}

\altaffiltext{1}{Cavendish Laboratory, University of Cambridge, 19 J. J. Thomson Ave., Cambridge CB3 0HE, UK}
\altaffiltext{2}{Kavli Institute for Cosmology, University of Cambridge, Madingley Road, Cambridge CB3 0HA, UK}
\email{sc888@mrao.cam.ac.uk}

\begin{abstract}

Himiko is one of the most luminous \lya\ emitters at $z=6.595$. It has three star forming clumps detected in the rest-frame UV, with a total SFR~=~20\sfr. We report the ALMA detection of the \cii 158$\mu$m line emission in this galaxy with a significance of 9$\sigma$. The total [CII] luminosity (\lcii = $1.2\times10^{8}$ \lsun)  is fully consistent with the local \lcii-SFR relation. The ALMA high-angular resolution reveals that the \cii\ emission is made of two distinct components. The brightest \cii\ clump is extended over 4~kpc and is located on the peak of the \lya\ nebula, which is spatially offset by 1 kpc relative to the brightest UV clump. The second \cii\ component is spatially unresolved (size $<$2~kpc) and coincident with one of the three UV clumps. While the latter component is consitent with the local \lcii-SFR relation,
the other components are scattered above and below the local relation. We shortly discuss the
possible origin of the
[CII] components and their relation with the star forming clumps traced
by the UV emission.
\end{abstract}


\section{Introduction} \label{sec:intro}

One of the key science goals of the Atacama Large Millimitre/submillimetre Array (ALMA) is the detection and investigation of star-forming galaxies  in the early Universe through the observations of the far-infrared (FIR) lines. In particular, ALMA observations of the strongest FIR lines, such as the \cii 158$\mu$m and \oiii 88$\mu$m, allow us to probe the properties of the interstellar medium (ISM) in galaxies at $z>6$ \citep[e.g.][]{Maiolino:2015,Pentericci:2016, Carniani:2017a,Inoue:2016,Matthee:2017,Smit:2017}, which are responsible for cosmic reionisation. 

The \cii\ (${^2}P_{3/2}\rightarrow{^2}P_{1/2}$) transition at 157.74 $\mu$m is one of the dominant coolants of the ISM.
 The \cii\ emission can be excited by the UV radiation field, produced
 by young stars,  in photodissociation regions (associated with
 neutral atomic and molecular gas), but is also associated with partly ionized gas. Therefore, this transition can be used as tracer of on-going star-formation
 in local and distant galaxies \citep[e.g.][]{De-Looze:2014, Vallini:2015,Herrera-Camus:2015}.
 
In the last five years ALMA observations targeting the \cii\ line emission in ``normal'' star-forming galaxies  $z>6$ (with  SFR~=~3-100~\sfr) have yielded a variety of results \citep{Ouchi:2013, Ota:2014, Willott:2015, Maiolino:2015, Pentericci:2016, Bradac:2016,  Knudsen:2016, Matthee:2017, Smit:2017}. The \lcii-SFR relation observed in these high-$z$ systems seems to have an intrinsic dispersion larger than that observed in the local Universe, which probably results from the wide range of global properties characterising primeval galaxies \citep[][Carniani et al. in prep.]{Vallini:2015, Matthee:2017, Lagache:2017}
The properties of the \cii\ emission are even more puzzling in luminous \lya\ emitters (LAEs), in which the \cii\ emission is very faint or not detected \citep{Ouchi:2013, Ota:2014,Willott:2015, Schaerer:2015}, placing these sources below the \lcii-SFR relation found in the local Universe \citep{De-Looze:2014}.  Simulations and models have attempted to explain this \cii\ deficit by taking into account the low metal content in high-$z$ galaxies, the excitation by the CMB radiation, and the effect of galactic expelling large amounts of gas  \citep{Vallini:2013, Vallini:2015, Pallottini:2017a, Pallottini:2017}. However, \cite{Matthee:2017} have recently reported the \cii\ detection in the most luminous  LAE at $z\sim7$, CR7, making the scenario even more complex. Indeed CR7 has a SFR=40 \sfr, which is comparable to the other previous LAEs showing a \cii\ deficit, but its \cii\ luminosity places this galaxy along the local SFR-\cii\ relation.

In this letter, we re-analyse archival ALMA \cii\ observations of Himiko, one of the most famous LAE, at  $z\sim6.595$ \citep{Ouchi:2009}. Himiko is located in the Ultra Deep South field. Its \lya\ nebula has a luminosity ${\rm L_{Ly\alpha}=4\times10^{43}}$ erg/s and extends over 17 kpc. The redshift of the source was spectroscopically confirmed trough  Keck/DEIMOS and VLT/Xshooter observations of the \lya\  \citep{Ouchi:2013,Zabl:2015}. Hubble Space Telescope (HST) observations with WFC3 in bands J$_{125}$ and H$_{160}$ (rest-frame UV at $z\sim6.595$) show that Himiko comprises of three bright clumps, A, B and C (Figure~\ref{fig:ciimap}) with similar UV magnitudes.
Clumps A and B are separated by a projected distance of 6.4 kpc, while the other two clumps have a separation of 2.5 kpc \citep{Zabl:2015}.  The \lya\  emission nebula embeds all the tree clumps, but its peak is located between  clumps A and B \citep{Ouchi:2013}. 
The first ALMA observations aimed at detecting \cii\ emission in Himiko system were obtained in Cycle 0 and presented in \cite{Ouchi:2013}, who reported a non detection for this galaxy placing an upper limit for the \cii\ luminosity at \lcii$<5.4\times10^{7}$ \lsun. Himiko was also observed with ALMA in Cycle 1 but the analysis of these new observations have not been published so far.

In this paper we present an analysis of both Cycle 0 and Cycle 1 data. When combined together, we show that such data present a clear
detection of \cii\, with a luminosity expected by the local \cii-SFR relation, and also present a clumpy morphology relative
to the UV distribution.

Throughout this paper we adopt the standard cosmological parameters $H_0 = 70$ km s$^{-1}$ Mpc$^{-1}$ , $\Omega_M = 0.30$, $\Omega_\Lambda$ = 0.70, according to which 1 arcsec at $z$ = 6.7 corresponds to a proper distance of 6.48 kpc. Astronomical coordinates and magnitude are given in the J2000 epoch and AB system, respectively.

\section{Observation and data reduction}

Himiko was observed with ALMA in band 6 in July 2012 and July 2015 as  part of the programs \#2011.0.00115.S (Cycle 0) and \#2012.1.00033.S (Cycle 1). The former program  was carried out by using a semi-compact array 
configuration with 25 12-m antennae and with a maximum baseline length  of 440 m. An extended-array configuration was instead used for  Cycle-1 observations, with 44 12-m antennae and baseline lengths up to 1600 m. The total on-source integration time was 3.2h for both programs. The perceptible water vapour ranged between 1.1 and 2.7 mm.  Both observations were performed  in frequency division mode with a total bandwidth of 7.5 GHz  and a spectral resolution of 0.488 kHz ($\sim$0.5 km/s).  One of the four spectral windows (each 1.875 GHz Wide) was set  to the expected frequency of the \cii\ line at $z=6.595$ ($\nu_{\rm obs}$=250.24 GHz). 

The absolute flux scale was established by observations of Callisto, Neptune, Ceres, and J0238+166. While 3C454.3, J2357-5311, J0423-0120, J2357-5311, and J0334-4008 were observed as bandpass calibrators. The phase calibration was based on the observations of J0423-013, J028-0047 and J2357-5311.

The two datasets have been calibrated by using the standard scripts for ALMA reduction corresponding to the release data of the observations. We used the  software CASA \citep{McMullin:2007} version 3.4.0 and 4.7.2  respectively for the two datasets. 
For each dataset we have generated the continuum map from the line-free channels and cubes with spectral channels of 5 km/s by using the CASA task {\sc clean} with a natural weighting. The resulting sensitivities and angular resolution of the continuum map and cubes are  listed in Table~\ref{tab:ALMA}. We note that the sensitivities and angular resolution of the final images obtained  from the Cycle-0 dataset are in agreement with those presented by \cite{Ouchi:2013}. 

We have also combined the observations of both programs to obtain maps and cubes with higher sensitivity. The imaging of the combined dataset has been performed by using CASA version 4.7.2  and natural weighting, achieving a continuum sensitivity of 9 $\mu$Jy/beam and an angular resolution of $\sim$0.30\arcsec\ at the observed frequency of \cii. 
To optimise the signal-to-nose ratio (S/N) of the \cii\ detection, we have produced a spatially smoothed cube  using a uv tapering of 0.2\arcsec\ so  to recover the spatially extended emission. We report the properties of 
combined and smoothed images  in Table~\ref{tab:ALMA}.

We have registered the ALMA and HST data by matching  ALMA calibrator  and foreground sources to the  GAIA Data Release 1 catalogue \citep{Gaia-Collaboration:2016}.

\begin{table*}
\small\begin{center}
\caption{ALMA data summary}\label{tab:ALMA}
\begin{tabular}{ccccccc }
\hline  
\hline
dataset  & t$_{\rm on}$ & antennae & beam & $\sigma_{\rm cont}$ & $\sigma_{\rm line}$  \\

 & [h] & & & [$\mu$Jy] & [$\mu$Jy] \\
  & (a) & (b) &(c)  & (d) & (e) \\
 \hline  \hline
2011.0.00115.S & 3.2 & 25 & 0.81\arcsec$\times$0.57\arcsec & 17 & 130 \\
2012.1.00033.S & 3.2 & 44 & 0.28\arcsec$\times$0.22\arcsec & 11 & 98 \\
\hline
combined & 6.4 & 44 & 0.39\arcsec$\times$0.31\arcsec & 9 & 88\\
combined and uv-tapered  & 6.4 & 44 & 0.48\arcsec$\times$0.40\arcsec & 11 & 101 \\
\hline  \hline
\multicolumn{7}{l}{%
  \begin{minipage}{10cm}%
 {\bf Notes}: (a) On-source exposure time. (b) Number of antennae. (c) Angular
 	resolution. (d) Continuum sensitivity at $\lambda _{rest}=$158 $\mu$m.
    (e) Sensitivity in a spectral channel of 100 km/s.
  \end{minipage}%
  }
\end{tabular}
\end{center}

\end{table*}

\section{Results}\label{sec:results}

\subsection{Continuum emission at 158$\mu$m}

The dust continuum emission at 158$\mu$m (${\rm S_{158\mu m}}$) is not detected, indicating a low dust mass content, as already discussed by \cite{Ouchi:2013}. We estimate a 3$\sigma$ upper limit on the continuum emission ${\rm S_{158\mu m}} <27 \ \mu$Jy that results into an upper limit on the infrared luminosity  ${\rm L_{IR}<2\times10^{10} \ L_{\odot}}$, if assuming grey-body emission
with a dust temperature of 35 K and a spectral emissivity index of 1.5 \citep{Ota:2014, Schaerer:2015, Willott:2015}.
If we assume the  ${\rm L_{IR}-SFR_{IR}}$ relation presented in \cite{Kennicutt:2012}, the infrared luminosity translates into an upper limit on the dust obscured ${\rm SFR_{IR}<4 \ M_{\odot}~yr^{-1}}$.
We note that such upper limit on ${\rm SFR_{IR}}$ is about four times lower than the total ${\rm SFR_{UV}}$ inferred from the rest-frame UV emission (see Table~\ref{tab:properties}), indicating that most of the star formation is unobscured by dust.

\subsection{[CII] line emission}\label{sec:cii}

\begin{figure*} \centering
{\bf (a)}
\includegraphics[width=1.28\columnwidth]{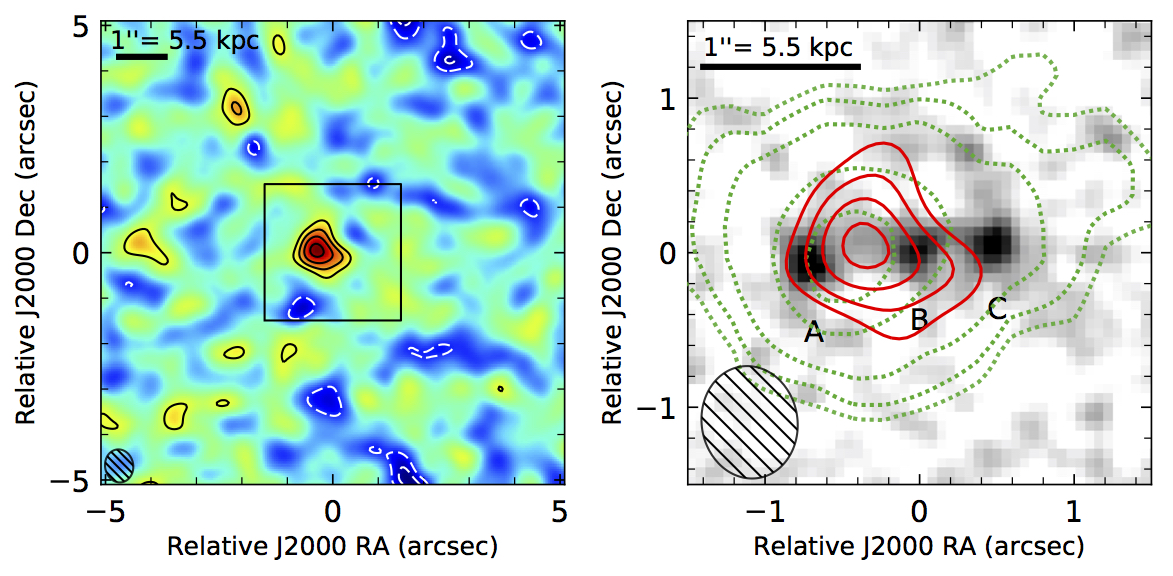}  
{\bf (b)}
\includegraphics[width=0.65\columnwidth]{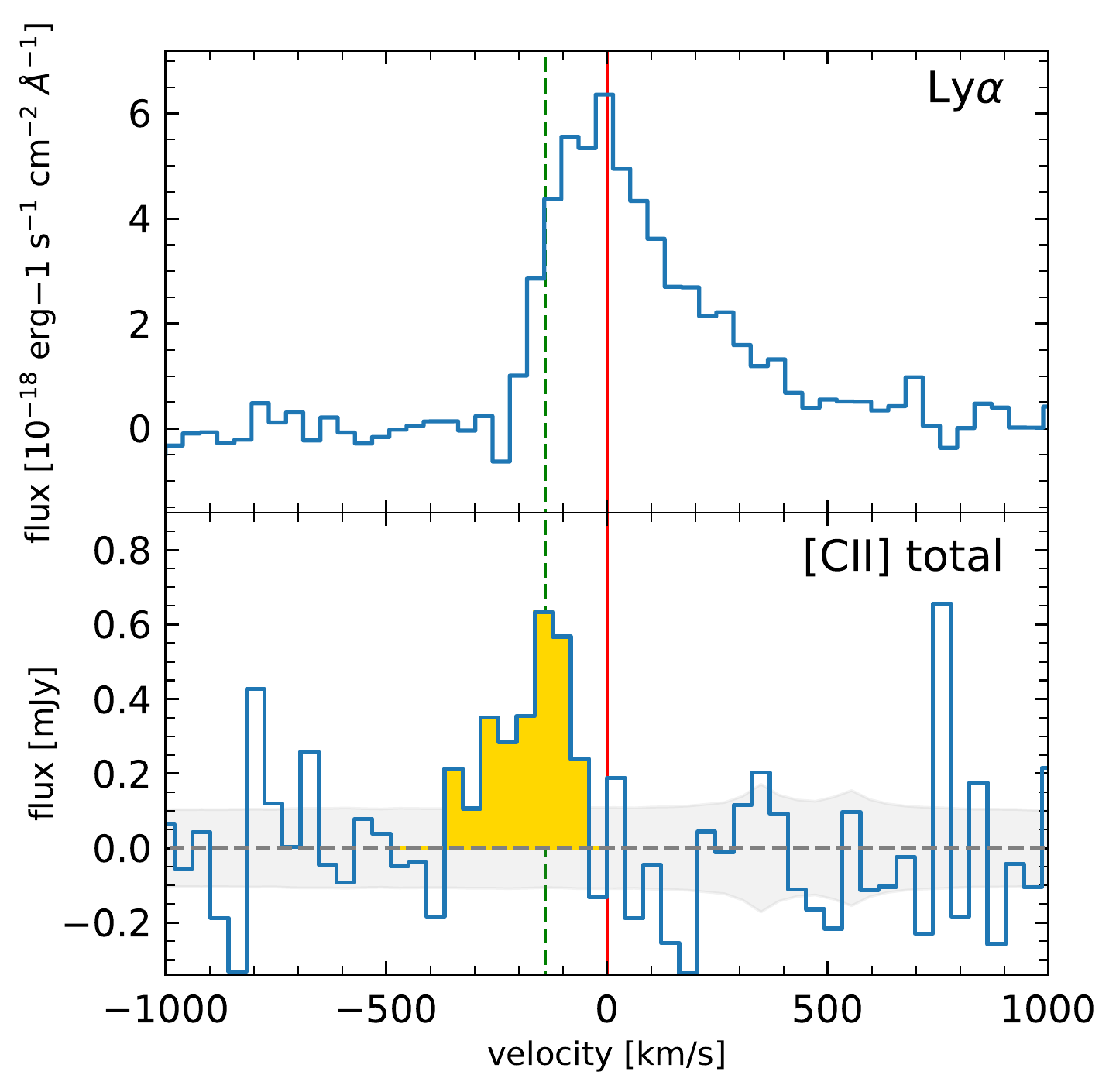}  \\
\caption{ {\bf(a)} The left panel shows the \cii\ map of Himiko obtained by integrating the uv-tapered cube over a velocity range between -320  and -50 km/s relative to the \lya\ redshift,  $z_{\rm Ly\alpha}=6.595$. Contours are shown in steps of ${\rm1\sigma=70\ \mu }$Jy~beam$^{-1}$ starting at $\pm2\sigma$ (white dashed contours are for negative fluxes). The zoom of the  central 3\arcsec$\times$3\arcsec\ around the location of the source is shown in the right panel, where the grey background image shows the HST/WFC3 J$_{125}$ observation, while the red contours indicate the  \cii\ emission with the same levels as in the left panel.
\lya\ intensity levels from \cite{Ouchi:2013} are indicated with green contours.
 The ALMA beam is indicated the in bottom-left corners of both panels. {\bf (b)} The top panel shows the \lya\ profile obtained from the X-shooter optical spectrum, while the \cii\ spectrum is shown in the bottom panel. The velocities are relative to the \lya\ redshift inferred by \cite{Ouchi:2009}. The dashed green line indicates the velocity inferred from the \cii\ profile. The grey shaded region indicates the 1$\sigma$ error as a function of velocity.}
\label{fig:ciimap}
\end{figure*}

We detect an emission line consistent with the \cii\ emission at the redshift and at the location of Himiko in the combined uv-tapered ALMA cube. Figure~\ref{fig:ciimap}a shows the map of the line extracted with a spectral width of 270 km/s and centred at -185 km/s relative to the redshift inferred by \lya, $z_{\rm Ly\alpha}=6.595$. The peak of the \cii\ emission has a significance of 5.5$\sigma$ and is located between the UV clumps A and B. However,
the \cii\ emission is spatially resolved, hence the significance of the total emission is higher than the peak flux in the map and specifically, as we will see below, the total significance is 9$\sigma$. The distribution of the \cii\ emission has a beam-deconvolved size of $(0.7\pm0.2)\arcsec\times(0.3\pm0.2)\arcsec$, which corresponds to a physical size of ${\rm (3.9\pm1.1) \ kpc \ \times(1.7\pm1.1) \ kpc}$, 
 embedding both UV clumps A and B.


The reliability of the \cii\ detection is also supported by the fact that the line is marginally detected at the same location in the two individual datasets. More specifically, in the Cycle-0 data the peak of the \cii\ emission is detected with a S/N=4.2 and  a flux level (${\rm S \Delta v = 112\pm26}$~mJy km/s) consistent with the value observed in the combined cube. The peak of the \cii\ emission is only marginally detected with S/N=3.2 (${\rm S\Delta v = 68\pm22}$ mJy km/s) in the Cycle-1 observations. The lower significance in the
Cycle 1 data is due to the high angular resolution of these data (7.6 and 3.1 times higher than that of the Cycle-0 and uv-tapered combined cube, respectively). Indeed,
although the Cycle-1 dataset has a sensitivity higher than that of Cycle-0, the diffuse emission is resolved out in the extended-array configuration observations and a fraction ($\sim40\%$) of the total \cii\ emission is missed. A similar scenario has already been reported by  \cite{Carniani:2017a} for the ALMA observations of the $z=7$ star-forming galaxy BDF-3299, where the bulk of the emission is ascribed diffuse gas extended on scales larger than 1 kpc and the new high-angular resolution observations have revealed only $\sim30\%$ of the total \cii\ emission. 

Since the emission is  resolved, we extract the \cii\ spectrum over a region within the $2\sigma$ \cii\ contours shown in Figure~\ref{fig:ciimap}a. The resulting spectrum, with a spectral rebinning of 40 km/s, is presented in Figure~\ref{fig:ciimap}b, together with  the optical spectrum of the \lya\ line obtained from X-shooter observations \citep{Zabl:2015}. 
The line is offset by $-145\pm15$ km/s relative to the $z_{\rm Ly\alpha}$ and has a line width of  $130\pm30$ km/s. The velocity offsets between \lya\ and \cii\ emission is similar to those estimated in other high-$z$ \cii-emitting sources observed with ALMA \citep{Maiolino:2015, Pentericci:2016, Matthee:2017}. 
While the \cii\ line is a good tracer of the systemic velocity of galaxies in the early Universe, the \lya\ profile is affected by intergalactic medium (IGM) absorption;  galactic outflows can also shift the peak of the line by up to few hundred km/s.
\cite{Zabl:2015} discuss that the strong blue asymmetry of the \lya\ profile is likely associated with the ISM/IGM absorption and, therefore, its centroid/peak results into an apparent redshift slightly higher than the systemic systemic redshift estimated from \cii, but fully consistent once the Ly$\alpha$ asymmetry is taken into account. 

The integrated intensity of the \cii\ line is  ${\rm S_{[CII]}\Delta v=108\pm12}$ mJy km/s. The resulting \cii\ luminosity   ${\rm L_{[CII]}=(1.2\pm0.2)}\times$10$^{8}$ \lsun, which is about two times higher than the upper limit estimated by \cite{Ouchi:2013}. This tension can
be explained with the fact that, not having {\it a priori} knowledge on the location of the \cii\ emission (spatially and spectrally)
\cite{Ouchi:2013} extracted an upper limit with an arbitrary line width and not taking into account fluctuations which may have been potentially
associated with a real signal. This issue may be common also to other \cii\ non-detections in the literature, in the sense that some of the
upper limits estimated in the past may be too low.

\begin{figure}
\centering
\includegraphics[width=1\columnwidth]{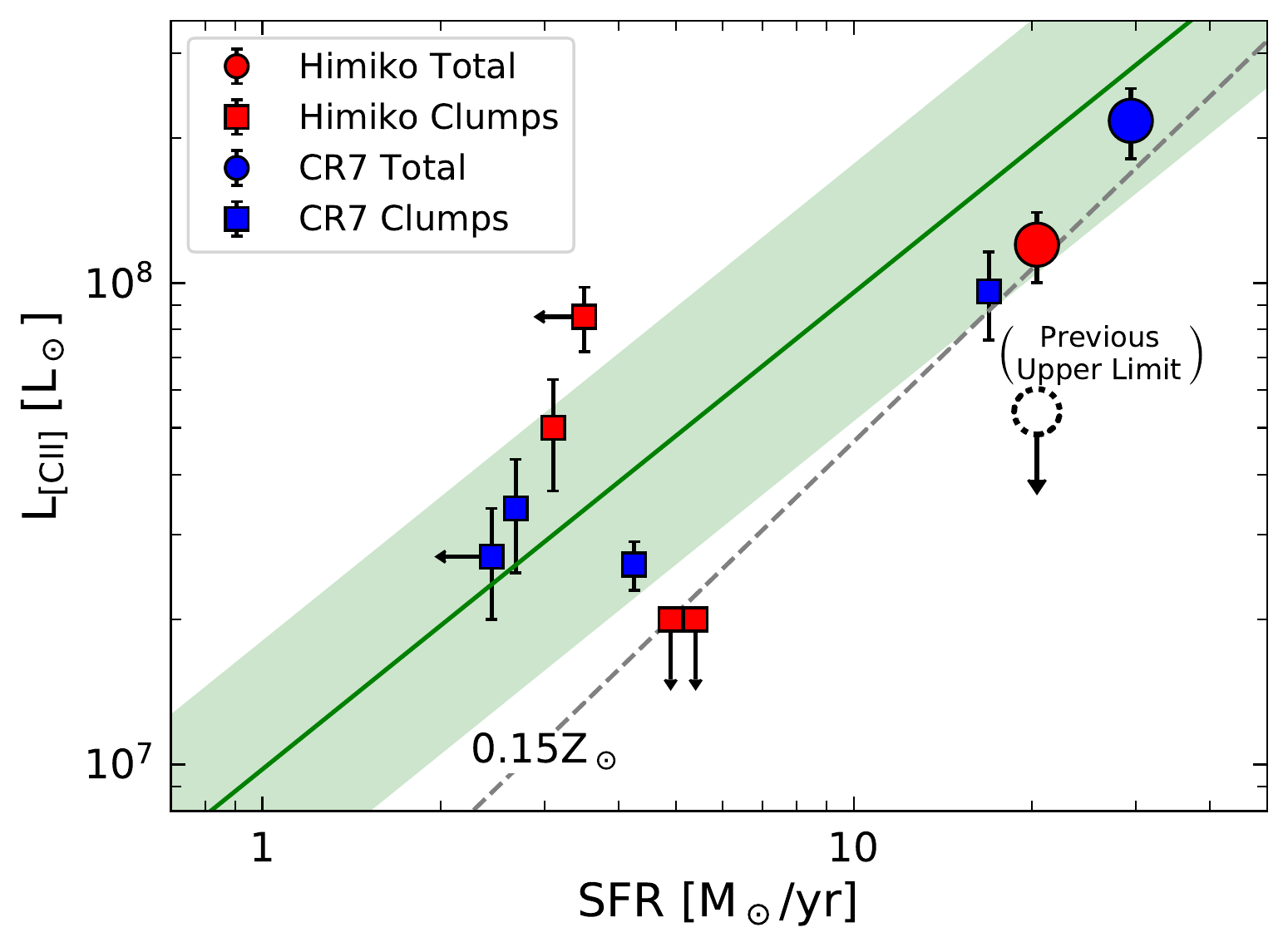}  \\
\caption{\lcii\ versus  SFR diagram. The green line shows the relations for local star-forming galaxies by \cite{De-Looze:2014}, while the dotted grey line is the best-fit relation for Z = 0.15Z$_\odot$ galaxies simulated by \cite{Vallini:2015}. The location of Himiko's total emission is shown with the large red circle. The location of Himiko's subclumps is shown with red squares. We also report the location
of CR7 and its subclumps. The dotted circle shows the previous upper limit on the \cii\ emission for Himiko obtained by \cite{Ouchi:2013}.}

 \label{fig:lcii_sfr}
\end{figure}

The measured \cii\ luminosity  is comparable to those measured in some other $z>6$ star-forming galaxies with similar SFR$_{\rm UV}$. In particular it is akin to the \lcii\ observed in the CR7 \citep{Matthee:2017}, which is a LAE with properties similar to Himiko (M$_{\rm UV}$ = -21.2, SFR = 44\sfr). In contrast to previous claims, our detection  places Himiko along the  local \lcii-SFR given by \cite{De-Looze:2014}, as illustrated in Fig.\ref{fig:lcii_sfr}.

Recent cosmological simulations by \cite{Vallini:2015} and \cite{Olsen:2017} show that at fixed SFR the \cii\ luminosity in galaxies decreases with the decreasing of the metallicity content. By using their best-fit relations we obtain a rough estimate of gas metallicity for Himiko of ${\rm Z\sim0.15Z_{\odot}}$, which is in line with that obtained from the spectral-energy-distribution fitting \citep[${\rm Z\sim0.2Z_{\odot}}$;][]{Zabl:2015}.  

\begin{table*}
\small
\begin{center}
\caption{UV and far-IR properties of Himiko}\label{tab:properties}
\begin{tabular}{lccccc }
\hline  
\hline
& \multicolumn{5 }{c}{Himiko}\\
 & Total & Clump A & \acii\ (\lya\ peak) & Clump B (\bcii) &  Clump C \\
 \hline
 RA [J2000] & - & 2:17:57.61 & 2:17:57.59 & 2:17:57.57 & 2:17:57.53 \\
 DEC [J2000] & - & -5:08:44.96 & -5:08:44.77 & -5:08:44.87 & -5:08:44.82 \\
 $z_{\rm Ly\alpha}$ & 6.595 & 6.595 & 6.595 & 6.595 & 6.595 \\
 ${\rm EW(Ly\alpha)}$ [\AA]  & $78^{+8}_{-6}$ & $68^{+14}_{-13}$ &  $>68$ &  $3^{+20}_{-18}$ & $6^{+12}_{-10}$ \\
 ${\rm J_{125}}$ & $24.99\pm0.08$  & $26.54\pm0.04$ & $>26.9$ & $27.03\pm0.07$ & $26.43\pm0.04$ \\  
 ${\rm H_{160}}$ & $24.99\pm0.10$ & $26.73\pm0.06$ & $>27.3$ & $27.04\pm0.08$ & $26.48\pm0.05$ \\
 ${\rm SFR_{UV} \ [M_{\odot}/yr]}$ & $20.4\pm1.5$ & $4.9\pm0.2$ & $<3.5$ & $3.1\pm0.2$ & $5.4\pm0.2$ \\
 \hline
 ${\rm S_{158\mu m} \ [\mu Jy]}$ & $<27$ & $<27$ &  $<27$ & $<27$ & $<27$ \\  
 ${\rm L_{IR} \ [L_{\odot}]}$ & $<2\times10^{10}$ & $<2\times10^{10}$ & $<2\times10^{10}$ & $<2\times10^{10}$ & $<2\times10^{10}$ \\ 
 ${\rm SFR_{IR} \ [M_{\odot}/yr]}$ & $<4$ & $<4$ & $<4$ & $<4$ \\
 $z{\rm_{[CII]}}$ & $6.5913\pm0.0004$ & - & $6.5906\pm0.0008$ & $6.5915\pm0.0003$ & - \\ 
 $z{\rm_{Ly\alpha}}-z{\rm_{[CII]}}$ [km/s] & $-145\pm15$ & - & $-175\pm30$ & $-140\pm10$  & - \\
 ${\rm FWHM_{[CII]} \ [km/s]}$ & $180\pm50$ & - &  $240\pm80$ & $70\pm20$ & - \\
${\rm S_{[CII]}\Delta v \ [mJy \ km/s]}$$^{\ddagger}$ & $108\pm12$ & $<26$  & $73\pm12$  & $44\pm12$ & $<26$  \\
 ${\rm L_{[CII]} \ [10^{8} \ L_{\odot}]}$$^{\ddagger}$ & $1.2\pm0.2$ & $<0.2$ & $0.85\pm0.13$ & $0.50\pm0.13$ & $<0.2$\\
\cii\ size [kpc] & $(3.9\pm1.1)\times(1.7\pm1.1)$ & - & $(3.7\pm0.6)\times(2.0\pm0.6)$ & $<2$ & -\\

 \hline  \hline
 \multicolumn{5}{l}{%
  \begin{minipage}{14cm}%
 {\bf Note}: 
 $\ddagger$ The upper limits on the \cii\ emission are estimated by assuming a $FWHM=100$ km/s.
  \end{minipage}%
  }
\end{tabular}
\end{center}

\end{table*}

\subsection{Multi-component system}\label{sec:clumps}

Given the multi-clump shape of Himiko in the rest-frame UV images and the extended \cii\ emission detected in the smoothed ALMA observations, in this section we investigate the morphology of \cii\ emission and its  connection with the UV and Ly$\alpha$ counterparts.

\begin{figure*}
\centering
\includegraphics[width=2\columnwidth]{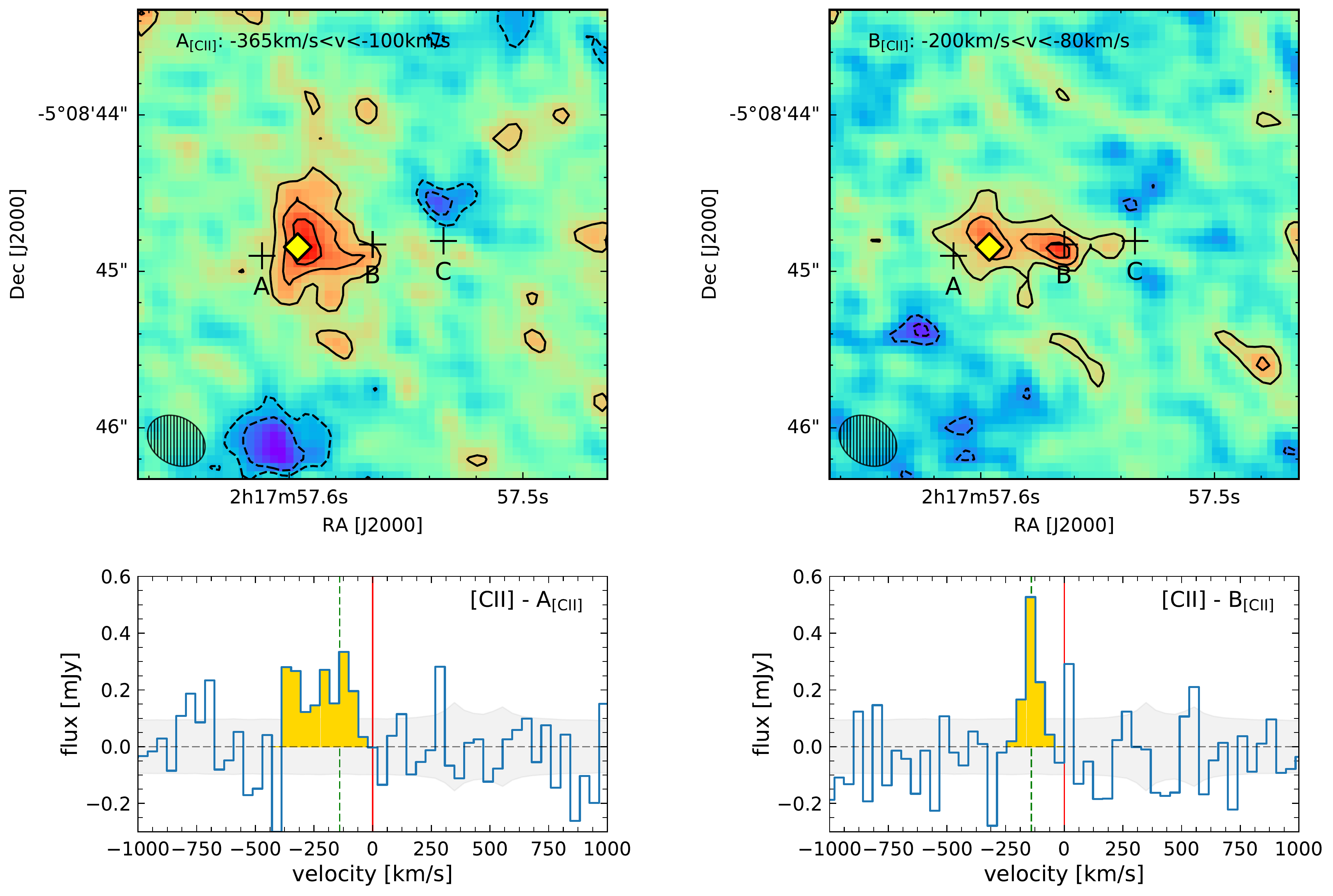}  \\

\caption{ The top panels show the flux maps of the two \cii\  components, clump \acii\ and \bcii. Contours show emission at levels of 2,3,4 and 5$\sigma$. The  position of the three UV clumps  is marked with black crosses, while the position of the \lya\ peak is indicated with a yellow diamond. The bottom panels show the \cii\ spectra of the two [CII] components.  Velocities are relative to the peak of the \lya\ peak (red line). The systemic velocity of the total \cii\  profile is indicated with a  red line.}
 \label{fig:clumps}
\end{figure*}

A channel map analysis  performed on the non-smoothed combined cube, which has an angular resolution of 0.39\arcsec$\times$0.31\arcsec, reveals that the extended \cii\ emission discussed in the previous Section is the result of two distinct components, which will be referred to as \acii\ and \bcii. The flux maps and spectra of these two components are shown in  Figure~\ref{fig:clumps}.

We identify the component \acii\ with a S/N=5 in the channel map  from -100 km/s to -365 km/s. The centroid of the emission is spatially offset by $\sim$0.2\arcsec\ toward the West relative to the UV position of the (brightest) clump-A, but it is fully consistent with the peak of the extended \lya\ nebula observed by \cite{Ouchi:2013} (yellow diamond in Fig.3). 
We exclude that this \cii\ component is directly associated with the UV clump A since the \cii\ emission level at the location of the UV emission is lower than $2\sigma$. As we will discuss later, positional offsets between UV and FIR line emission are not so rare in primeval galaxies \citep{Maiolino:2015,Willott:2015,Carniani:2017a}.
By fitting a 2D elliptical Gaussian profile to the channel map, we find  that the \cii\ emission of component \acii\ is spatially resolved with a beam-deconvolved size of ${\rm(0.68\pm0.10)\arcsec\times(0.41\pm0.10)\arcsec}$, which corresponds to a physical size of $\sim3.7$ kpc.  
The \cii\ profile peaks at $-175\pm30$ km/s and has a line width of $240\pm80$ km/s. We measure \lcii$=(0.85\pm0.13)\times10^{8}$ \lsun\ that corresponds to the $\sim$70\% of the total \cii\ luminosity inferred for Himiko.

The other component, \bcii, peaks at $-175\pm30$ km/s and is narrower ( FWHM = $70\pm20$ km/s) than the former component. The knot of the [CII] emission is co-aligned with the UV position of clump-B, indicating that [CII] and UV emission arise from the same region. Since  components \bcii\ and \acii\ overlap in velocity,  the flux map \bcii\ shows also a tail emission extending to the west that is assocaited with the \acii\ component. The core of the \bcii\ emission is not spatially resolved indicating the the \cii\ line is powered by a compact source with physical size $<2$ kpc. For this component we infer a \cii\ luminosity \lcii$=(0.50\pm0.13)\times10^{8}$ \lsun.

Since around the \lya\ redshift we do not detect any  emission close to the  UV clumps A and C, we search for line emission in the ALMA cube between -1000 km/s and 1000 km/s relative to the  redshift of \lya\ and with level of significance $>3\sigma$. 
Only a putative detection is found at $\sim-500$ km/s and located at the UV position of the clump C. Because of the low significance (S/N=3.2), the emission can be spurious due to noise fluctuation, hence we consider it as a non detection.
For the two UV clumps A and C, we therefore infer an upper limit on the \cii\ luminosity    \lcii$<0.2\times10^8$ \lsun\ where we assume a line width of 100 km/s.

We note that while the \cii\ emission in clump $\rm B(=B_{[CII]})$ is fully consistent with the local \lcii --SFR relation, clumps A, C and \acii\ 
are scattered outside the local relation (Figure~\ref{fig:lcii_sfr}), as observed in  other high-$z$ galaxies \citep[e.g.][]{Maiolino:2015}.

As discussed in \cite{Carniani:2017a},
positional offsets between UV and \cii\ emission may be ascribed to spatially distinct regions of the galaxy, and of their circumgalactic environment, characterised by different physical properties. For instance, the low \cii\ luminosity at the location of the UV regions can be interpreted as a consequence of  a local low metal enrichment level in  these star-forming regions, but also in terms of strong feedback
ionizing or expelling gas  \citep{Vallini:2015,Katz:2016, Olsen:2017}. 
Spatially-offset \cii\ emission may also be explained  in terms of dust obscuration and/or outflowing/inflowing gas. In the former scenario  \cii\ is excited in-situ by star-formation whose UV emission is heavily dust-obscured. In this context we note that \cite{Ouchi:2013} infer a dust attenuation for Himiko of E(B-V)=0.15, which can hide a significant fraction of star forming regions traced by the UV emission.
On the other hand, the spatially-offset \cii\   can be associated to a satellite clumps
in the process of accreting, or clumps expelled by galactic outflows; in these cases the \cii\ emission is excited by the UV radiation of the closest star-formation region (e.g. clump A). This last scenario is also supported by the fact that  Himiko reveals a triple  major merger event whose extended \lya\ nebula emission may be  powered by both star formation and galactic winds  \citep{Ouchi:2013,Zabl:2015}. 

In summary UV and \cii\ emission can trace different regions that should be treated as different sub-components of the same system. A detailed discussion of multiple sub-components observed in $z>5$ star-forming galaxies, as well as their offset relative to the star forming regions traced
by the UV emission, is presented in a companion paper (Carniani et al. in prep.).

\section{Conclusions}

We have presented the analysis of archival ALMA data targeting the \cii\ emission in the famous and luminous LAE Himiko, at $z=6.595$. We have detected the FIR line with a level of significance of 9$\sigma$ for the total, spatially integrated emission.

The measured luminosity of the line (\lcii$=1.2\times10^{8}$ \lsun) is fully consistent with the local \lcii-SFR relation,
thereby mitigating the discrepancy with the local \lcii-SFR relation claimed in previous studies.  

The ALMA data reveal that the \cii\ profile is blueshifted by -145 km/s relative to the peak of \lya, consistent with the fact that the asymmetric,
blueward profile of \lya\ is associated with absorption by the neutral intergalactic medium. 

The \cii\ emission is spatially resolved over $\sim4$ kpc and breaks into two sub-components. The location of the faintest \cii\ component (which is spatially
unresolved) is consistent with the UV emission of  clump B, which is the central UV clump out of the three associated with Himiko. Instead, the brightest \cii\ component is spatially extended (over 4~kpc) and its centroid is coincident with the peak of the \lya\ nebula, and is spatially offset by $\sim0.2\arcsec$ ($=$1kpc) relative to the nearest (and brightest) UV clump. While the \cii\ luminosity of clump B is fully consistent with the local \lcii-SFR relation, all other clumps are scattered around the local relation, both above and below.

As already discussed in other works, the offsets between \cii\ and UV emission, and the multi-clump nature of the \cii\ emission, are likely associated with various effects, such as dust obscuration (preventing the detection of UV associated with \cii\ emission), strong feedback (ionizing or removing  gas
in intense star forming regions), minor/major mergers and circumgalactic accreting/outflowing gas. A more extensive discussion of
the multicomponent \cii\ properties of high-$z$ galaxies, their morphologies and offsets relative to the UV emission, is presented in a
more extensive paper, discussing several more galaxies at $z>5$, in a companion paper (Carniani et al. in prep.).

Finally, we note that the previous works had not detected the \cii\ line in Himiko, giving an upper limit a factor of two lower than
our detection. This is due to the difficulty of properly estimating the upper limit without prior knowledge about the location, redshift
and width of the \cii\ emission. This issue may affect also other past results; in particular, past upper limits on the \cii\ emission in other
high-z galaxies may be too low.

\acknowledgements Acknowledgements. This paper makes use of the following ALMA data: ADS/JAO.ALMA\#2011.0.00115.S  and  ADS/JAO.ALMA\#2012.1.00033.S; which can be retrieved from the ALMA data archive: https://almascience.eso.org/ alma-data/archive. ALMA is a partnership of ESO (representing its member states), NSF (USA) and NINS (Japan), together with NRC (Canada) and NSC and ASIAA (Taiwan), in cooperation with the Republic of Chile. The Joint ALMA Observatory is operated by ESO, AUI/NRAO and NAOJ. SC acknowledges support by the Science and Technology Facilities Council (STFC). RM and RA  acknowledge support by the Science and Technology Facilities Council (STFC) and the ERC Advanced Grant 695671 ``QUENCH''.

\bibliographystyle{aasjournal}
\bibliography{bibliography_himiko.bib}
\end{document}